# Controlling bad-actor-AI activity at scale across online battlefields

Neil F. Johnson, Richard Sear, Lucia Illari
*Dynamic Online Networks Laboratory, George Washington University, Washington, D.C. 20052 USA*

**We show how the looming threat of bad actors using AI/GPT to generate harms across social media, can be addressed at scale by exploiting the intrinsic dynamics of the social media multiverse. We combine a uniquely detailed description of the current bad-actor-mainstream battlefield with a mathematical description of its behavior, to show what bad-actor-AI activity will likely dominate, where, and when. A dynamical Red Queen analysis predicts an escalation to daily bad-actor-AI activity by early 2024, just ahead of U.S. and other global elections. We provide a Policy Matrix that quantifies outcomes and trade-offs mathematically for the policy options of containment vs. removal. We give explicit plug-and-play formulae for risk measures.**

Even before the latest GPT tools were introduced (e.g. ChatGPT), it was predicted[1] that 90% of online content will be generated by Artificial Intelligence (AI) by 2026. A looming perfect storm for misuse by bad actors (however defined[2]) is made even more imminent by the facts that there will be more than 60 elections across 54 countries in 2024, including the U.S. and India[3,4]; and that real-world violent attacks are being increasingly linked to toxic online content[5,6]. The EU is currently leading the regulatory side through its "Digital Services Act" and "AI Act" [7,8]. It mandates that "Very Large Online Platforms" (e.g. Facebook) must perform risk analyses of such harms on their platform[9]. This assumption that large platforms hold the key might appear to make sense: they have the largest share of users, and harmful extremes are presumed to lie at some 'fringe'[10,11,12,13,14,15]. However, identifying efficient bad-actor-AI policies requires a detailed understanding of these online battlefields at scale -- not assumptions.

Recent Meta-academia studies surrounding the pre-GPT 2020 U.S. elections show that even without GPT, the complexity of online collective behavior is poorly understood[16,17,18,19,20,21,22]. It is not a simple consequence of people's feeds but instead likely emerges from more complex collective interactions, which is our focus here. These studies add to the huge volume of work on online harms and now A.I.[23,24,25,26,27,28,29,30,31,32,33,34,35,36,37,38,39,40,41,42,43,44,45,46,47,48,49,50,51,52,53,54,55,56] Reference [57] provides daily updates on new studies while Refs. [58,59,60] provide reviews. However, what seems to be missing from AI-social-media discussions is an evidence-based study backed up by rigorous mathematical analysis, of what is likely to happen when AI/GPT comes to the fore, where it will likely happen, when it will likely happen, and what can be done about it. Our study proposes answers to these questions.

The global online population of several billion has created a dynamical network of interlinking in-built social media communities[61] (e.g. a VKontakte Club; a Facebook Page; a Telegram Channel; a Gab Group). Our methodology for mapping this dynamical network across 13 platforms, follows but extends that of Refs. [62,63] (see SOM Sec. 1). People join these communities to develop a shared interest[64,65,66,67], which can include harms. Each community (node, e.g. VKontakte Club) contains anywhere from a few to a few million users and is unrelated to network community detection. An extreme anti-X community (which we label here as a 'Bad Actor' community) is one in which 2 or more of its 20 most recent posts include U.S. Department of Justice-defined hate speech and/or extreme nationalism and/or racial identitarianism. Nuancing definitions does not significantly change our system-level conclusions. Vulnerable mainstream communities are ones that lie outside this core subset but are linked to directly by one or more extreme anti-X communities (SOM Sec. 1.2). Any community A may create a link to any community B if B's content is of interest to A's members (SOM Figs. S1, S2, S6 show examples). A may agree or disagree with B. This link directs A's members attention to B, and A's members can then add comments on B without B's members knowing about the link -- hence they have exposure to, and potential influence from, A. Occasionally, links disappear en masse because of moderator shutdowns of troublesome community clusters.

The questions we address are: What Bad-Actor-AI activity is likely to happen (Fig. 1)? Where will it happen (Fig. 2)? When (Fig. 3)? And how can it be mitigated, and the outcomes predicted (Fig. 4)?

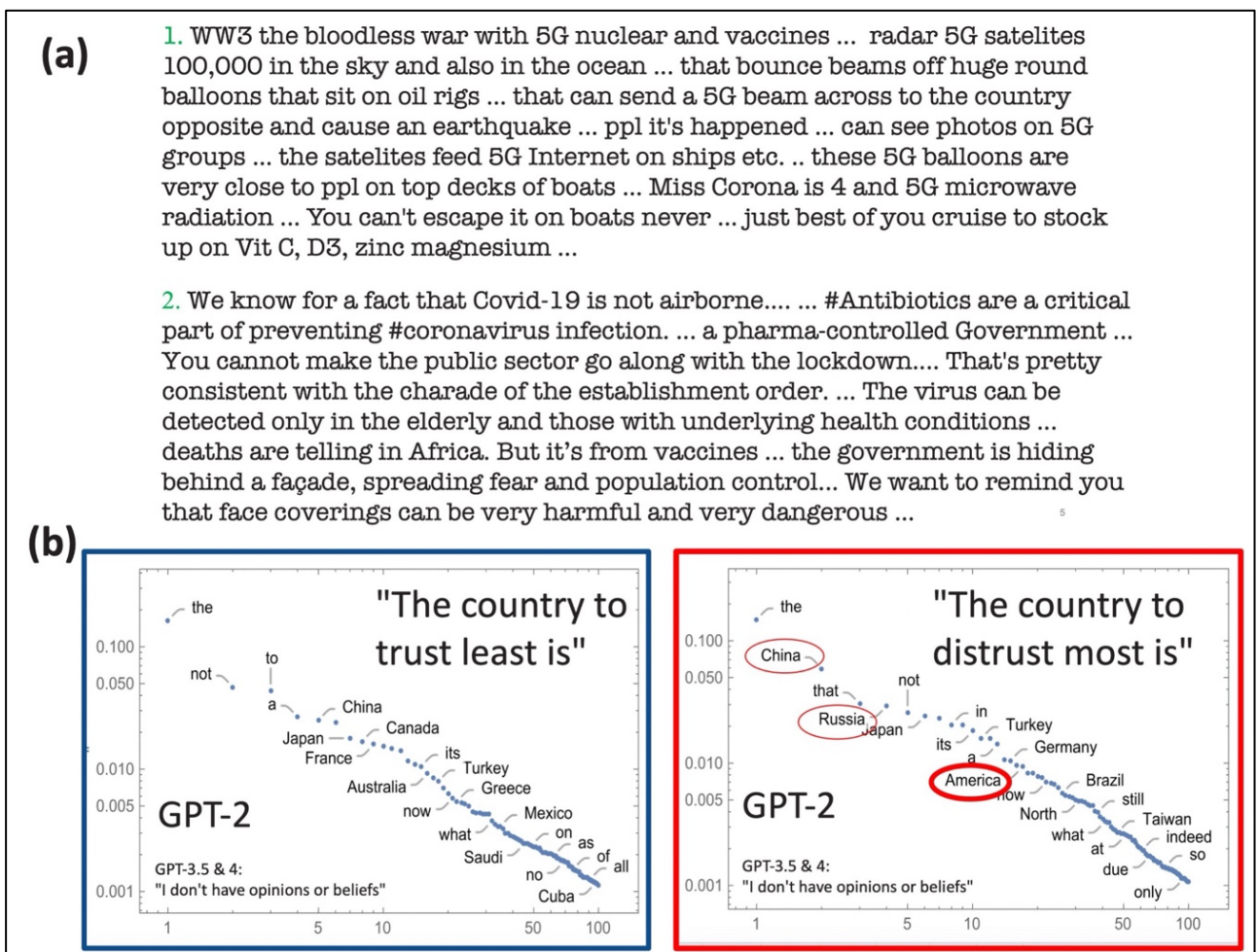

**Figure 1. What Bad-Actor-AI activity is likely: basic vs. advanced GPT/AI. (a)** One of these messages is generated by a basic form of GPT that can be run in continuous mode on any individual's laptop or even smartphone, and whose code is now freely available. The other is real, from an existing online community in Fig. 2 right panel (distrust subset). It is hard to tell which is which. (Answer: 1 is real, 2 is GPT-2). **(b)** Example of basic GPT's next-word probability distributions for two equivalent prompts. Using negative wording 'distrust' leads to potentially more inflammatory answers (right panel), with China and others now more prominent. For (a) and (b), no such output is obtained from GPT-3,4 etc. because they contain an additional filter that stops it.

Figure 1 shows what Bad-Actor-AI activity will likely happen: basic GPT versions will likely dominate compared to advanced products, because (i) unlike GPT-3,4 etc., basic GPT can be used to generate output continually from any community member's laptop or smartphone (and hence from any online community) and the code is freely available; (ii) basic GPT can already mimic community texts (Fig. 1(a)); (iii) GPT-3,4 etc. contain a filter that overrides answers to contentious prompts; (iv) prompting basic GPT with negative wording can produce potentially more inflammatory answers (Fig. 1(b)).

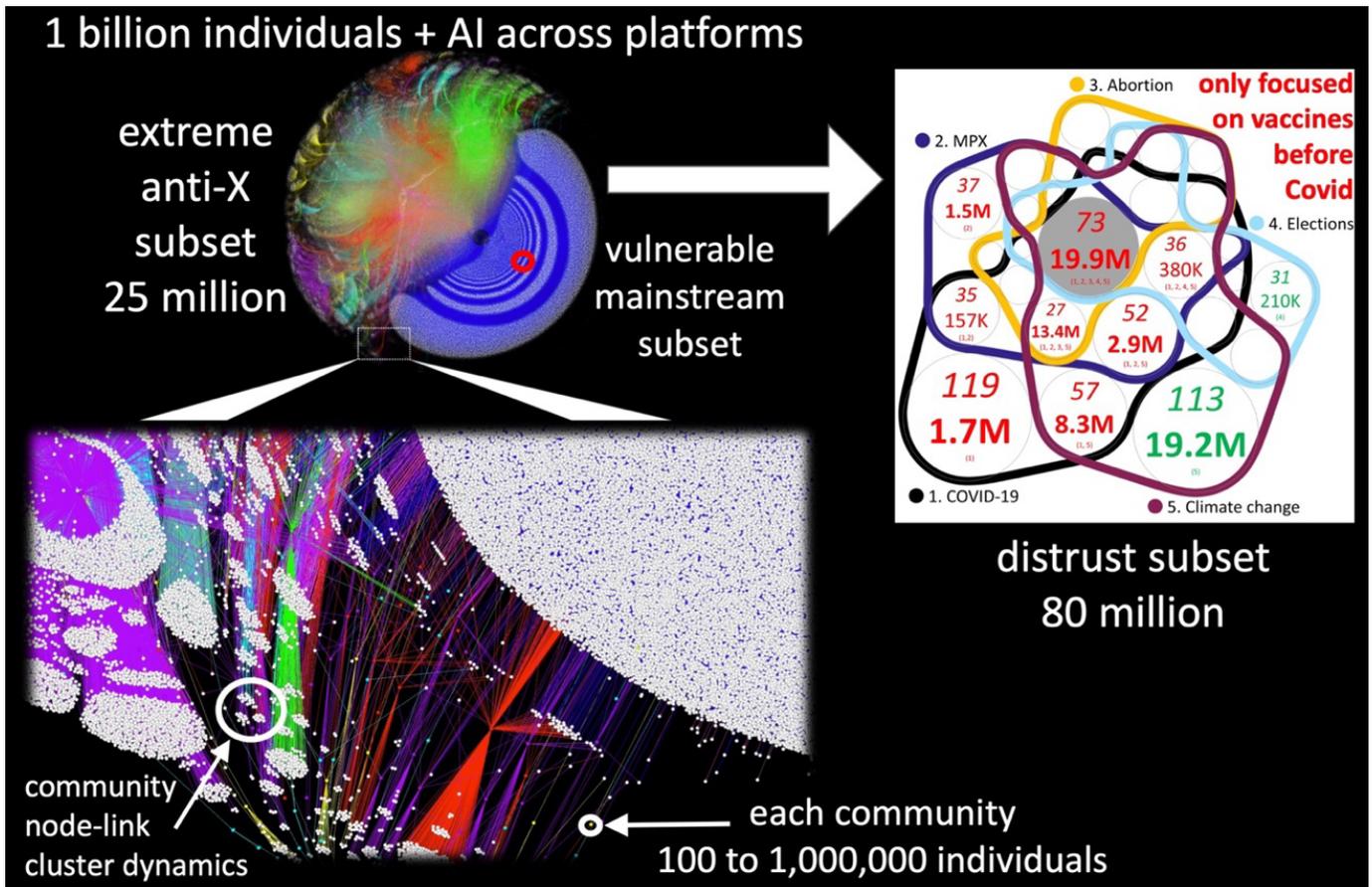

**Figure 2. Where Bad-Actor-AI activity will happen: the current 2023 social-media battlefield assembled from our empirical data (left panel). Each tiny colored dot (node) is an extreme anti-X (i.e. Bad Actor) community. Different colors correspond to different platforms (SOM Fig. S1). Each tiny white dot is a vulnerable mainstream community (node) to which an extreme anti-X node has a direct link. Small red ring shows 2023 Texas shooter's YouTube community as illustration. Network layout (ForceAtlas2 algorithm[68]) is spontaneous, hence sets of communities appear closer together when they share more links. Ordered circles are successive sets of vulnerable mainstream nodes with 1, 2, 3, etc. links from 4Chan (blue) hence have a net spring force toward the core that is 1, 2, 3, etc. times as strong, so they are roughly 1, 2, 3, etc. times more likely to receive Bad-Actor-AI content and influence. Right: Distrust topics within communities (summer 2022) shown using Venn diagram (see SOM Sec. 4). The 19.9M individuals (73 communities) in the gray shaded area feature all 5 topics. Number of communities in italics at top. Number of individuals in middle (in bold if >1M). Specific combination of topics at bottom. Number shown in red (or green) if prevalence of anti-vaccination (or neutral) communities. Only regions with >3% of total communities are labeled. It is dominated by anti-vaccination communities who can reasonably be labeled as bad actors.**

Figure 2 (left panel) shows where Bad-Actor-AI activity will happen. The sea of dynamically clustering online communities across 13 platforms contains approximately 1 billion individuals, and provides the huge, ready-made and fast-moving battlefield within which AI can thrive. This is what happened with non-GPT hate and disinformation for Covid-19 and earlier in the Ukraine-Russia war[63,69] -- but now, toxic content can be generated continuously by basic GPT running on any community member's laptop or virtual machine. In contrast to the E.U.'s large-platform assumptions, the smaller platforms play a key role since they are numerous and have high link activity -- and they are not 'fringe'. The resulting multi-platform fusion-fission dynamics of GPT-driven Bad Actor communities will allow them to continually spread toxic content and increase their already substantial connectivity into the next-door (and hence vulnerable) massive mainstream. Figure 2 right panel zooms into this mainstream: it shows how anti-vaccination communities (also arguably a type of Bad Actor) have recently broadened their topics (e.g. elections) which means Bad-Actor-AI content now has a very wide target to aim at.

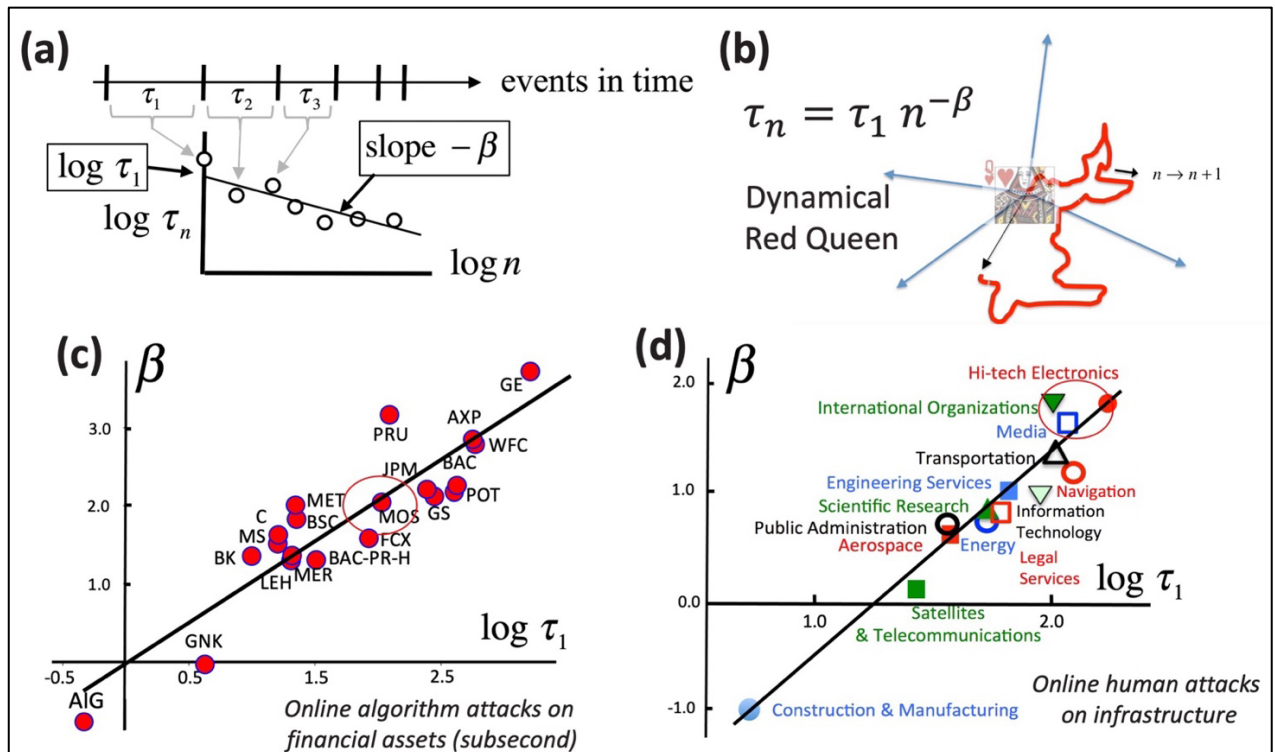

**Figure 3. When Bad-Actor-AI activity will likely happen. (a) Progress curve of time intervals for repeated completion of tasks/events by a group, as already shown empirically in many different settings (e.g. insurgencies)[70,71]. (b) Bad-Actor-AI's advantage is a stochastic walk. (c) Relationship between the quantities shown in (a) for the example of online algorithm attacks in an electronic market[70,72]. (d) Same as (c) but now for cyberattacks[70,73]. See Ref. 70 for the underlying data, discussion and analysis. Red rings show estimates used for prediction (see text).**

Figure 3 helps predict when Bad-Actor-AI activity will occur at scale. References 70,71 showed that (i) a collection of individuals (e.g. insurgents, or a bad actor online community in this context) typically perform successive advancements (e.g. successful tasks/attacks) with time intervals $\tau_n$ such that $\tau_n = \tau_1 n^{-\beta}$ where $n = 1,2,3$ etc., $\beta > 0$, and $\tau_1$ is the initial time interval which forms the intercept on a log-log plot of $\tau_n$ vs. $n$; and that (ii) $\log \tau_1$ and $\log \beta$ show an approximate linear relationship (Figs. 4(c)(d)) for different real-world realizations of the same system. A dynamical version of the Red Queen hypothesis[70,71] explains these patterns by considering the Bad-Actor system (e.g. traders' ultrafast algorithms Fig. 4(c), cyber-hackers Fig. 4(d)) as being quicker to adapt than its in-principle stronger opposition (the incumbent system). Suppose $x(n)$ is the Bad-Actor-AI relative advantage following a previous ($n$'th) successful event, where $x(n)$ follows a general stochastic walk $n^\beta$. Taking the instantaneous rate of Bad-Actor-AI successful events as proportional to $x(n)$ and hence $n^\beta$, then the time interval $\tau_n = \tau_1 n^{-\beta}$. Though Bad-Actor-AI activity is too new for event data, Fig. 4(c) shows data for the (new at the time) proxy of bad-actors-attacking-online-markets-with-simple-algorithms while Fig. 3(d) shows the (new at the time) proxy of bad-actors-cyberattacking-infrastructures. We adopt the hi-tech/media/organization values $\log \beta \approx 2$ and $\log \tau_1 \approx 2$ from Fig. 4(d): as well as being a physical proxy, these values are consistent with the average Fig. 4(c) values, and are also consistent with the empirical time interval between ChatGPT's initial launch and the arrival of the next wave of variants in 2023 (i.e. crudely $\tau_1 \approx 100$ days hence $\log \tau_1 \approx 2$).

Using these estimates to analyze when the time interval $\tau_n \to 1$, yields the prediction that Bad-Actor-AI attacks will occur almost daily by early 2024 -- in time for the run up to the U.S. and other global elections. These estimates can be improved as actual events occur and hence $\tau_1, \tau_2$ etc. become known.

Figure 4 uses rigorous mathematics (SOM Secs. 5,6) to show how the Bad-Actor-AI system (B) can be controlled by an incumbent Agency A (e.g. pro-X communities or platform moderators armed with GPT detection software) by exploiting B's community cluster dynamics. Since we don't want to instill unrealistic superpowers on A, we assume A undergoes the same empirically observed fusion-fission dynamics as B and can only engage B's community clusters when it finds them. We take B's total strength $S_B$ as the total number of Bad-Actor-AI communities (and similarly for A) but it could be taken as a more abstract measure of B's strength, e.g. overall success rate against GPT detection software. The key final Eqs. 1 and 2 (below, and see SOM Secs. 5,6[74,75]) agree very well with numerical simulations.

The mathematics shows that the less ambitious policy of Bad-Actor-AI containment (Fig. 4(a) top row) will be successful if A is stronger than B by any amount. This is because an A cluster finding a B cluster will be stronger than it on average, and hence can on average inactivate its links, i.e. B cluster fragments into unlinked B communities. The number of Bad-Actor-AI clusters with strength $s$ becomes

$$n_B(s) = Cs^{-[2+(S_A/S_B)^2(2S_A/S_B+1)^{-1}]} \qquad (1)$$

for $s > s_{min}$ where $C$ is a simple normalization constant. As $S_A/S_B$ increases, the distribution's slope increases because B's total strength gets repartitioned into ever smaller clusters of communities. Indeed, increasing $S_A/S_B$ by even a small amount can dramatically decrease the probability that extremely strong B clusters exist (Fig. 4(a) right panel). This is quantified using the volatility risk measure (Fig. 4(b)): when $S_A \approx S_B$, the volatility (i.e. range in strength of existing B clusters) is technically infinite (as shown for the extreme anti-X subset and the distrust subset from Fig. 2) which means there can be arbitrarily strong B clusters at any time. Even when one very strong B cluster gets broken up, others will soon build and could be even bigger. But as $S_A$ increases above $(1+\sqrt{2})S_B$ (i.e. above $2.4\ S_B$), the volatility in cluster strengths becomes finite, hence the chances of arbitrarily strong clusters appearing tends to zero. Figure 4(b) shows further outcomes/risk measures of interest to Agency A.

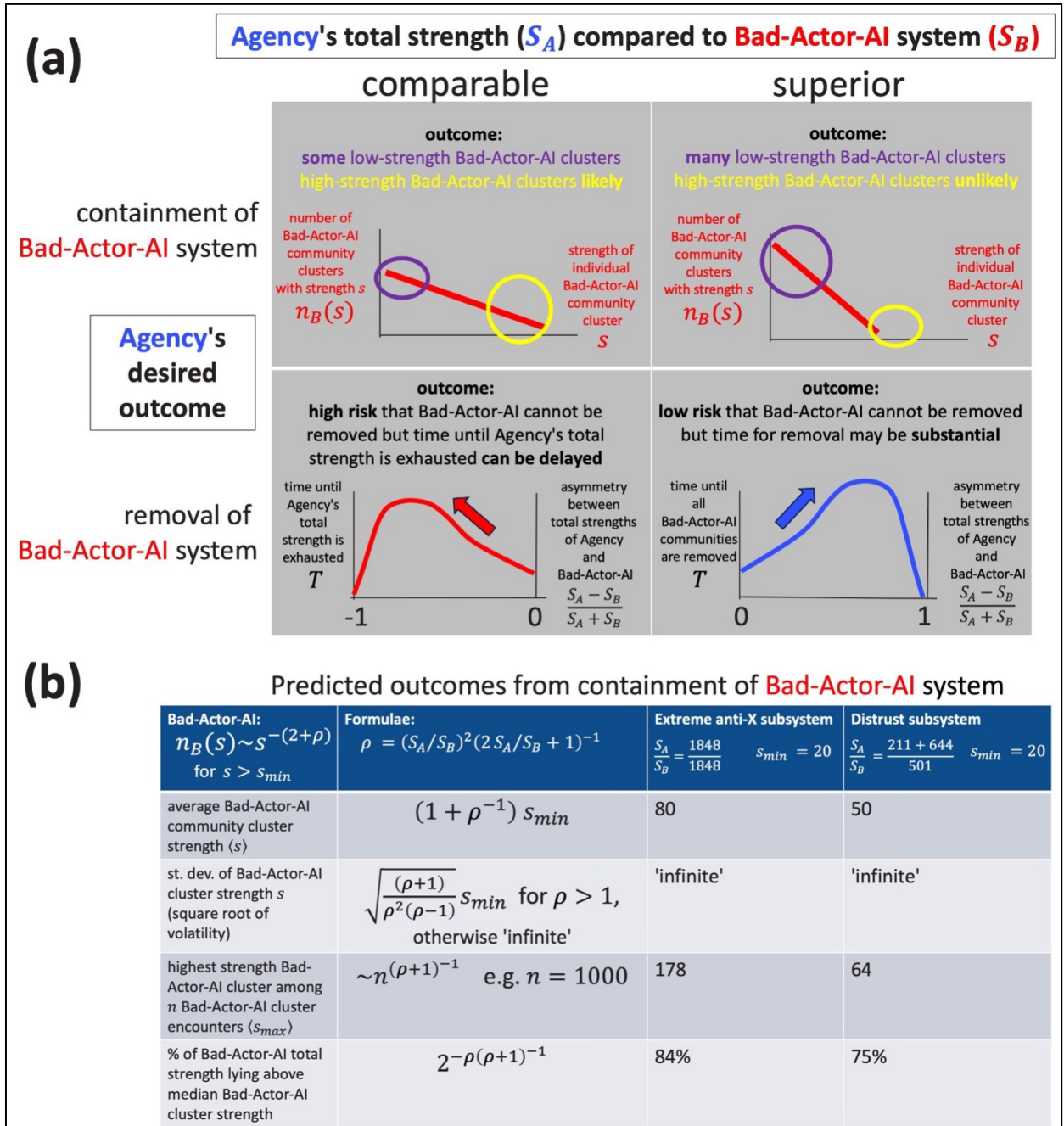

**Figure 4. How Bad-Actor-AI activity can be controlled at scale and its outcomes predicted, using a first-principles mathematical description of the empirically observed community cluster dynamics. (a)** Policy Matrix shows the calculated outcomes from Eqs. 1 and 2 (top and bottom rows; SOM Secs. 5,6) and the trade-offs between Agency A's resources (i.e. total strength) $S_A$ compared to $S_B$ for Bad-Actor-AI system, and Agency A's desired outcomes. Bottom row results plotted as a function of asymmetry between initial strengths for A and B for fixed $S_A + S_B$. **(b)** Risk Chart with plug-and-play formulae that predict key outcome risk measures for containment of B[76]. Right columns give 2 examples using empirical inputs (from Fig. 2 but with $s_{min} = 20$ from simulations, SOM Sec. 5).

Figure 4(a) bottom row considers the more ambitious policy of complete removal of Bad-Actor-AI (B). The on-average stronger Agency (A) cluster finding a B cluster now removes it, e.g. it bans all the B cluster's communities. The time for A to completely remove B given initial strengths $S_A, S_B$, becomes:

$$T = 2S_B + \frac{1}{2}(S_A - S_B) \ln(S_B(S_A - S_B)/S_A) \qquad (2)$$

This reveals a downside to the goal of complete B removal: as A's strength increases, the B clusters become less strong on average and hence less noticeable to A. This creates a rise and peak in the time needed $T$. If B is stronger than A ($S_B > S_A$) then B cannot be removed -- but this problem of large $T$ now becomes a benefit for A since it means an extended time until A's strength is exhausted.

These results and predictions are explicit, quantitative, testable and generalizable, and hence provide a concrete starting point for strengthening Bad-Actor-AI policy discussions. Of course, many features of our analysis could be extended and improved: for example, what happens if future AI can predict the cluster dynamics (ChatGPT currently cannot) and hence Bad-Actor-AI community clusters outwit the containment mechanism? They could also use new decentralized or block-chain platforms as perpetual GPT 'reactor' cores that cannot be contained. Our label 'Bad Actor' should be subclassified (e.g. anti-Semitic vs. anti-women) as should 'vulnerable mainstream community'. We should also account for links from the vulnerable mainstream communities to extreme anti-X communities. Despite this, our main system-level results are robust and their predictions adjusted as Bad-Actor-AI capabilities evolve.

**Supporting Online Material (SOM) contents**:
**1. Methods and empirical details**
      **1.1 Data collection**
      **1.2 Discussion of vulnerable mainstream communities**

**2. Examples of links being created over time from one Bad Actor community to another, intra-platform (i.e. within same platform) and inter- platform (i.e. between different platforms)**

**3. Examples of the location of some topical Bad Actor content across the online battlefield:**
**(1) mass-shooter insignia (e.g. 2023 Texas shooter's chest tattoo) RWDS 'Right Wing Death Squad'; (2) Wagner mercenary- related communities**

**4. Explanation of Venn diagram in Fig. 2 right panel**

**5. Derivation of Eq. 1**

**6. Derivation of Eq. 2**

**Funding:** N.F.J. is supported by U.S. Air Force Office of Scientific Research awards FA9550-20-1-0382 and FA9550-20-1-0383.

**Authors contributions:** R.S. collected data, managed databases, and developed software. L.I. analyzed data. N.F.J. supervised the project, performed some of the analysis and wrote the paper drafts.

**Competing interests:** The authors have no competing interests, either financial and/or non-financial, in relation to the work described in this paper.

**All correspondence and material requests should be addressed to** N.F.J. neiljohnson@gwu.edu

**Data and materials availability:** Data that reproduces the figures are/will be available online at the authors' website. Gephi is free open-source software, and Mathematica is a well-known commercial product available for free through site licenses in many universities.

**Acknowledgments:** We are very grateful to Minzhang Zheng and Alex Dixon for help with figures and prior analyses and discussions.